\documentclass[twocolumn,english,aps,prl,floatfix,amssymb,superscriptaddress]{revtex4}
\usepackage[T1]{fontenc}
\usepackage[latin9]{inputenc}
\usepackage{color}
\usepackage{amsmath}
\usepackage{amssymb}
\usepackage{graphicx}
\usepackage{epstopdf}
\usepackage{soul}

\makeatletter
\@ifundefined{textcolor}{}
{%
 \definecolor{BLACK}{gray}{0}
 \definecolor{WHITE}{gray}{1}
 \definecolor{RED}{rgb}{1,0,0}
 \definecolor{GREEN}{rgb}{0,1,0}
 \definecolor{BLUE}{rgb}{0,0,1}
 \definecolor{CYAN}{cmyk}{1,0,0,0}
 \definecolor{MAGENTA}{cmyk}{0,1,0,0}
 \definecolor{YELLOW}{cmyk}{0,0,1,0}
}

\usepackage{babel}
\usepackage[caption=false]{subfig}
\makeatother

\usepackage{babel}
\begin{document}

\title{Inducing time reversal invariant topological superconductivity and
fermion parity pumping in quantum wires}

\author{Anna Keselman}

\affiliation{Department of Condensed Matter Physics, Weizmann Institute of Science,
Rehovot, Israel 76100}

\author{Liang Fu}

\affiliation{Department of Physics, Massachusetts Institute of Technology, Cambridge,
MA 02139, USA}

\author{Ady Stern}

\affiliation{Department of Condensed Matter Physics, Weizmann Institute of Science,
Rehovot, Israel 76100}

\author{Erez Berg}

\affiliation{Department of Condensed Matter Physics, Weizmann Institute of Science,
Rehovot, Israel 76100}

\date{\today}

\begin{abstract}
We propose a setup to realize time-reversal invariant topological
superconductors in quantum wires, proximity coupled to conventional
superconductors. We consider a model of quantum wire with strong spin-orbit
coupling and proximity coupling to two s-wave superconductors. When
the relative phase between the two superconductors is $\phi=\pi$
a Kramers' pair of Majorana zero modes appears at each edge of the
wire. We study the robustness of the phase in presence of both time-reversal
invariant and time-reversal breaking perturbations. In addition, we
show that the system forms a natural realization of a fermion
parity pump, switching the local fermion parity of both edges when
the relative phase between the superconductors is changed adiabatically
by $2\pi$.
\end{abstract}

\maketitle

\emph{Introduction.}$-$Over the last few decades, it has been realized
that there is a deep and unexpected relation between the properties
of matter and topology. At zero temperature, there exist phases of
matter that are distinguished by an underlying topological structure
encoded in their ground-state wave functions. These phases are often
characterized by a finite energy gap in their bulk, and protected
gapless edge states with unusual properties. The bulk can either be
insulating, as in the case of the recently discovered topological
insulators (TI)\cite{Hasan2010,Qi2011}, or superconducting\cite{Read2000,Schnyder2008,Ryu2010a}.
Phases of the latter type, known as ``topological superconductors''
(TSC), support anomalous zero-energy Andreev edge states which are
robust as long as the bulk quasi-particle gap remains open. These
edge states have attracted much attention due to their possible future applications for topologically protected
quantum information processing\cite{Kitaev2003}. Recently, it has
been predicted that the one-dimensional variant of a TSC can be realized
by proximity-coupling a semiconducting quantum wire to a superconductor
(SC)\cite{Kitaev2001,sau2010generic,Lutchyn2010,oreg2010helical,Cook2011}.
The resulting TSC phase has particle-hole symmetric modes at  zero energy, localized at the edges of the wire, known as Majorana zero modes\cite{Majorana-review}.
Signatures of such zero modes have been observed in recent experiments\cite{Mourik2012,Das2012,Rokhinson2012}.

In the presence of time-reversal invariance (TRI), different types
of TSC can arise\cite{Qi2009,Qi2010}. These phases are solid-state
analogues of the B phase of superfluid $^{3}$He\cite{Leggett1975,VolovikBook}.
The alloys Cu$_{x}$Bi$_{2}$Se$_{3}$\cite{Hor2010,Fu2010,Wray2010,Sasaki2011,Hsieh2012,Kirzhner2012}
and Sn$_{1-x}$In$_{x}$Te\cite{Sasaki2012} are possible candidates
for these phases. In addition, it has been proposed that proximity-coupling
an unconventional superconductor
to a quantum wire can stabilize a one dimensional TSC phase which
supports a \emph{Kramers pair} of Majorana zero modes at its edge\cite{Wong2012,Zhang2012,Nakosai2013},
protected by TRI. This edge modes are characterized by an anomalous
relation between the fermion parity of the edge and time-reversal
symmetry\cite{Qi2009}.

\begin{figure}
\subfloat[\label{fig:setup}]{\includegraphics[scale=0.4]{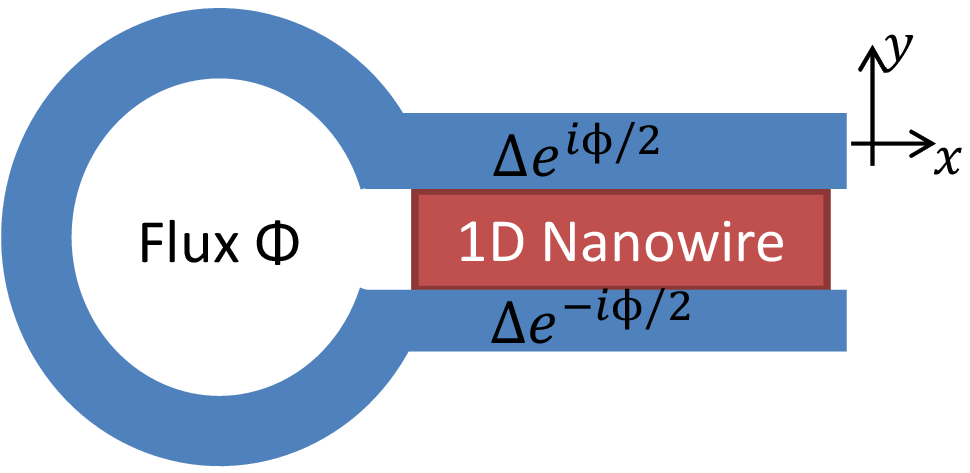}}
\subfloat[\label{fig:two-wires}]{\includegraphics[scale=0.4]{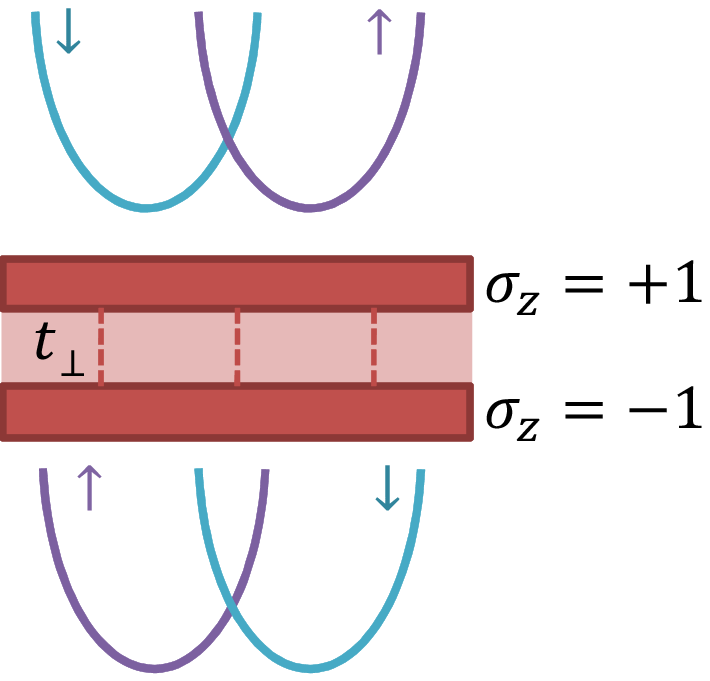}}

\caption{
(a) The general setup proposed for the realization of a TRI TSC. The
TRI topological phase is obtained for $\phi=\pi$.
(b) Specific model considered for the nanowire.
The spin orbit coupling on the two chains comprising the wire
has an opposite sign as indicated by the energy dispersion curves. A finite tunneling
amplitude $t_{\perp}$ between the two chains opens a gap at the crossing points.
}
\end{figure}

In this Letter, we propose a different setup to realize TRI TSC in
quantum wires. The setup is shown schematically in Fig. \ref{fig:setup}. A quantum
wire with strong spin-orbit coupling is proximity coupled to two s-wave
superconductors from either side\cite{Nakosai2012-note}. We assume that the relative phase
$\phi$ between the two superconductors can be controlled externally,
e.g. by connecting the two superconducting leads and threading a flux
$\Phi$ through the resulting superconducting loop. The relative phase
is then $\phi=2\pi\Phi/\Phi_{0}$, where $\Phi_{0}=hc/2e$ is the
superconducting flux quantum. We assume that the flux is applied far
away from the wire, such that the magnetic field in the region of
the wire is zero. The system has TRI for $\phi=0$ and $\phi=\pi$.
For $\phi=0$, we expect the induced SC state in the wire to be topologically
trivial. In contrast, for $\phi=\pi$ we show that a TRI TSC state
is formed in the wire under a broad range of circumstances, and a
Kramers' pair of Majorana states appears at each edge of the wire. This follows from a general criterion for TRI TSC in centrosymmetric systems\cite{Sato2009, Fu2010}, which we extend to one-dimensional systems below.

Unlike the previous proposals\cite{Wong2012,Zhang2012,Nakosai2013}, the setup we present requires only coupling to \emph{conventional} superconductors, and may thus be easier to realize. The key for achieving TRI TSC in our system is that the induced pairing potential in the quantum wire is odd under spatial parity, due to the $\pi$ phase difference between the two external superconductors.

In addition, we consider the effect of time-reversal breaking perturbations,
such as a deviation of the relative phase from $\pi$ or a Zeeman
field. These perturbations split the degeneracy of the edge states.
Surprisingly, however, we find that the topological character of the
system is not completely lost. Instead, the system forms a natural
realization of a \emph{fermion parity pump}, switching the local fermion
parity as well as flipping the local spin density at both edges when $\phi$ is changed adiabatically by $2\pi$. This is a generalization of the adiabatic charge pump proposed by Thouless\cite{Thouless1983}.

\emph{Conditions for TRI TSC.}$-$We consider a system which has TRI
and particle-hole symmetry, such that $\mathcal{T}^{2}=-1$, $\mathcal{C}^{2}=1$,
where $\mathcal{T},\mathcal{C}$ are the time reversal and particle-hole
operators, respectively (class DIII\cite{Altland1997}).
The phases in such systems are classified by a $\mathbb{Z}_{2}$ invariant
in spatial dimensions $d=1,2$ and by a $\mathbb{Z}$ (integer) invariant
in $d=3$. A sufficient condition for TRI TSC in centrosymmetric systems
in $d=2$ and $3$ was derived in Refs. \cite{Sato2009,Fu2010,Qi2012}.
The condition states that if $(1)$ the pairing is odd under inversion
and opens a full SC gap and $(2)$ the number of TRI momenta enclosed
by the Fermi surface in the normal (non-SC) state is odd, then the
system is in a TSC state. We have extended the condition
to 1D systems \cite{SM}, for which $\left(2\right)$
above is replaced by the requirement that in the normal state the number of
(spin-degenerate) Fermi points between $k=0$ and $k=\pi$ is odd.
In 1D centrosymmetric systems, the condition is both sufficient and necessary.

Applying this condition to the setup of Fig. \ref{fig:setup}, we
see that for $\phi=\pi$ the induced pairing is odd under a spatial
inversion $\vec{r}\rightarrow-\vec{r}$, which interchanges
the two superconductors. Suppose that the wire is made of a material
with a centrosymmetric crystal structure. Then, if the number of spin-degenerate
bands crossing the Fermi level of the wire is odd, and if the bulk
of the wire is fully gapped by the proximity effect, then the resulting
state is necessarily a TRI TSC.

Note that, although our condition makes no reference to the necessity
of spin-orbit coupling (SOC) in the wire, SOC is essential to realize
a TSC \cite{SM}. Therefore, we expect
that in the absence of SOC the bulk remains gapless for $\phi=\pi$,
invalidating one of the requirements for TSC.

\emph{Model.}$-$As an illustration, we consider a simple model of
a centrosymmetric quantum wire with SOC. Our model consists of two
coupled chains, with a SOC term originating from Rashba nearest neighbor
hopping and consistent with inversion symmetry. The Hamiltonian is given by
\begin{equation}
\mathcal{H}=\sum_{k}\psi_{k}^{\dagger}H_{k}\psi_{k}^{\vphantom{\dagger}}.\label{eq:H}
\end{equation}
Here, $\psi_{k}^{\dagger}=\left(c_{k}^{\dagger},-is_{y}c_{-k}\right)$
is a spinor in Nambu space, where $c_{k}^{T}=\left(c_{1\uparrow k},c_{1,\downarrow,k},c_{2,\uparrow,k},c_{2,\downarrow,k}\right)$,
and $c_{lsk}^{\dagger}$ creates an electron with momentum $k$ and
spin $s$ at chain $l=1,2$. We use $\vec{s}$ to denote Pauli matrices
in spin space and $\sigma_{z}=\pm1$ for the upper/lower chain,
respectively; see Fig. \ref{fig:two-wires}. The Hamiltonian matrix
is written as $H_{k}=H_{0k}\tau_{z}+H_{\Delta}\tau_{x}$, where $\vec{\tau}$
are Pauli matrices that act on the Nambu (particle-hole) space, and
the matrices $H_{0k}$, $H_{\Delta}$ are given by
\begin{eqnarray}
H_{0k} & = & \xi_{k}+\lambda_{k}s_{z}\sigma_{z}-t_{\perp}\sigma_{x},\\
H_{\Delta} & = & \Delta\sigma_{z},\label{eq:Hdelta}
\end{eqnarray}
where $\xi_{k}=2t\left(1-\cos k\right)-\mu$ and $\lambda_{k}=2\lambda\sin k$.
The parameters $t$ and $t_{\perp}$ are nearest neighbour hopping
amplitudes along the chains and between the chains respectively, $\lambda$ is the SOC strength, and $\mu$ is the chemical potential.
$H_{\Delta}$ describes the proximity coupling to two superconductors
with opposite phases. Inversion symmetry is implemented by the operator
$\mathcal{P}=\sigma_{x}$, that interchanges the two chains,
followed by $k\rightarrow-k$.

The Hamiltonian (\ref{eq:H}) can be diagonalized by a Bogoliubov transformation.
The spectrum is given by
\begin{equation}
E\left(k\right)=\pm\left[\xi_{k}^{2}+\lambda_{k}^{2}+t_{\perp}^{2}+\Delta^{2}\pm2\sqrt{\xi_{k}^{2}t_{\perp}^{2}+\xi_{k}^{2}\lambda_{k}^{2}+t_{\perp}^{2}\Delta^{2}}\right]^{1/2}.\label{eq:Ek}
\end{equation}
Each band is doubly degenerate, as expected from the symmetry of the
system under time reversal and inversion.
For $\left|t_{\perp}\right|>\left|\mu\right|$ and $\Delta=0$, there is a single spin-degenerate band crossing the Fermi level.
For $0<\left|\Delta\right|\ll t_{\perp}$, the spectrum becomes
fully gapped with a minimum gap $\Delta_{\mathrm{min}}\approx\left|\Delta\right||\lambda_{k_{F}}|/\sqrt{t_{\perp}^{2}+\lambda_{k_{F}}^{2}}$
at the Fermi points. ($k_{F}$ is the Fermi momentum.) In this case,
the condition above is satisfied, and the system is in the TRI TSC phase.
The gap remains open as long as $t_{\perp}^{2}>\mu^{2}+\Delta^{2}$.
For $\lambda=0$ the system remains gapless.

\begin{figure}
\subfloat[\label{fig:spectrum_B=0_TwoWires}]
{\includegraphics[scale=0.16]{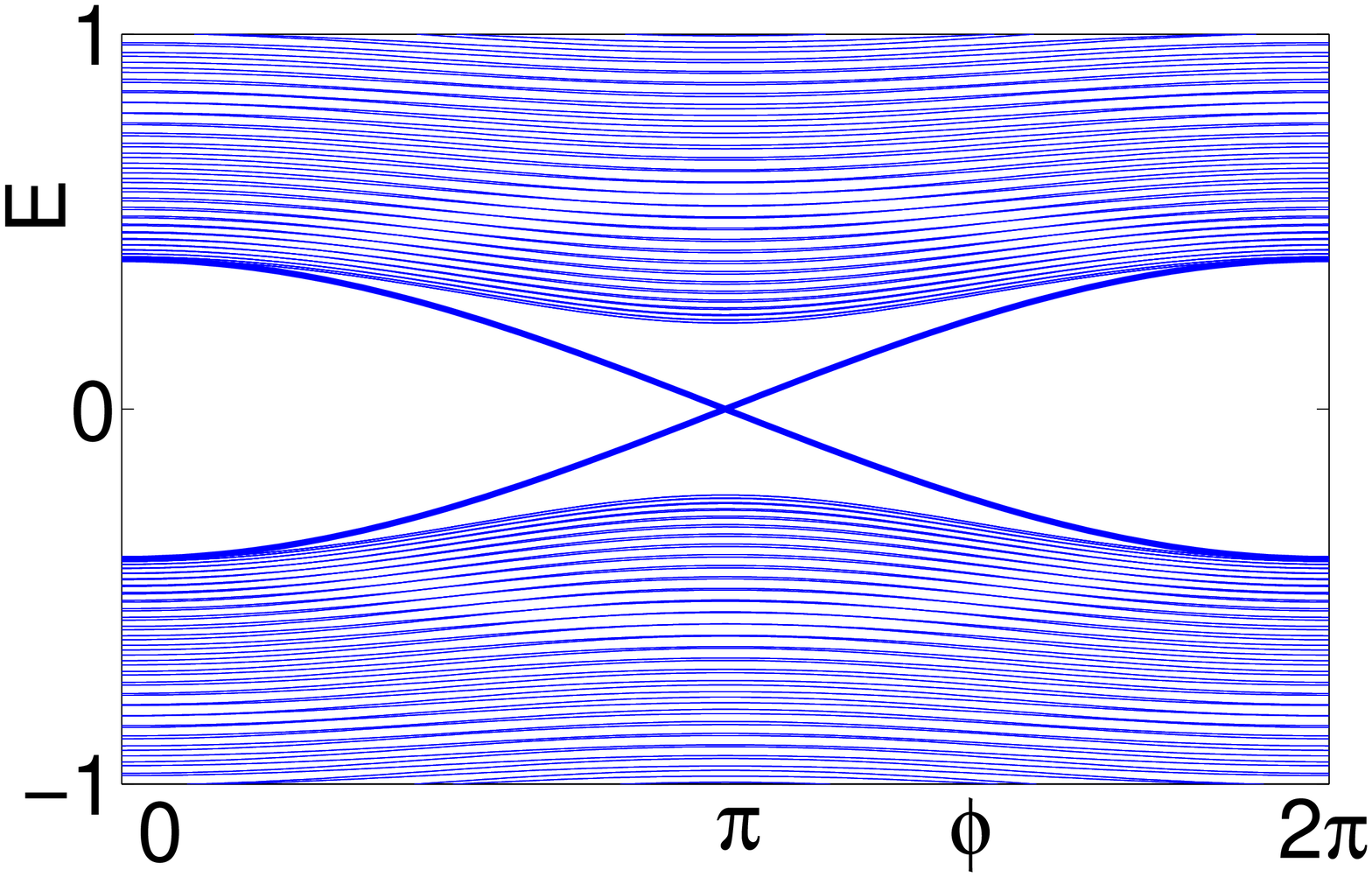}}
\subfloat[\label{fig:spectrum_B!=0_TwoWires}]
{\includegraphics[scale=0.16]{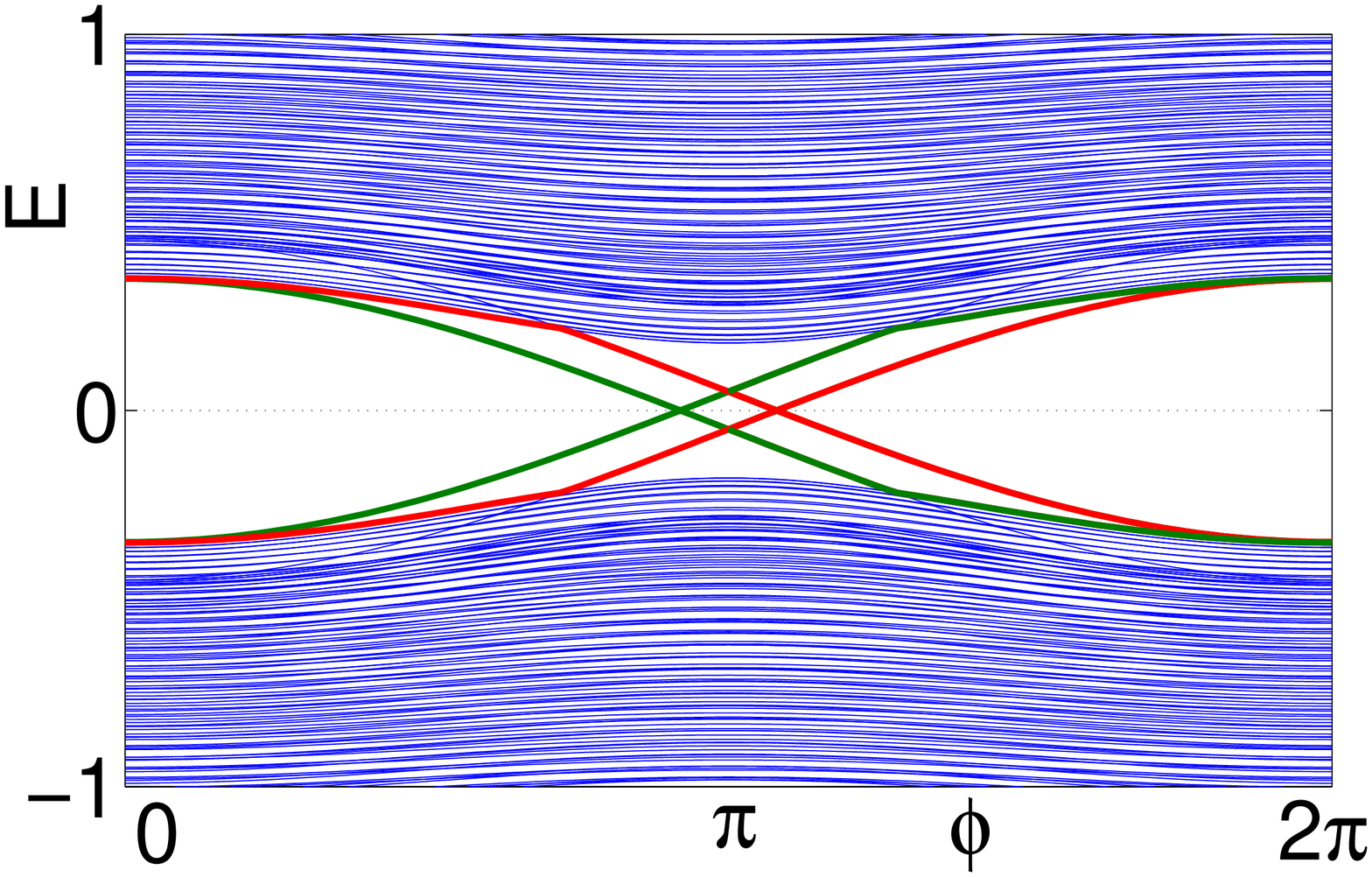}}

\caption{Energy spectrum of a finite nanowire as a function of the relative
phase between the two superconductors for (a) zero and (b) non zero
magnetic field. The length of the nanowire is $N_{x}=200$. The lattice
model parameters are $t_{\perp}=2.5,\ t=1,\ \lambda=1$. The chemical
potential is set to $\mu=0$ and the magnitude of the SC pairing is
$\Delta=0.4$. (a) For $\phi=\pi$ the system is TRI and in the topological
phase as can be seen from the presence of zero-energy states. (b)
Here $B=0.05$ in the $z$ direction. Energy levels plotted in red,
green correspond to edge states on the opposite edges of the wire.
At $\phi=\pi$ the two Majorana modes on each edge split. The crossing
at zero energy does not disappear, but is shifted away from $\phi=\pi$.}
\end{figure}

At the edge of a system in the TRI TSC phase, we expect to find a
single Kramers' pair of Majorana zero modes. To see that this is indeed
the case, we note that the model (\ref{eq:H}) can be thought of as
two copies of the model considered in Refs.\cite{sau2010generic,oreg2010helical},
\begin{equation}
\tilde{H}_{k}=\left(\xi_{k}+\tilde{\lambda}_{k}\sigma_{z}\right)\tau_{z}-B\sigma_{x}+\Delta\tau_{x},\label{eq:H_yuval}
\end{equation}
describing a semi-conducting wire with Rashba spin-orbit coupling
in an external magnetic field given by $B=t_{\perp}$. Note that in
(\ref{eq:H}), $s_{z}$ is conserved, and can be replaced by its eigenvalue
$\pm1$. Then, the unitary transformation $U=e^{i\frac{\pi}{4}\left(1-\sigma_{z}\right)\left(1-\tau_{z}\right)}$
maps $H_{k}$ to $\tilde{H}_{k}$ with $\tilde{\lambda}_{k}=s_{z}\lambda_{k}$.
The model (\ref{eq:H_yuval}) has been shown\cite{sau2010generic,oreg2010helical}
to support a single Majorana zero mode at the edge for $t_{\perp}^{2}>\mu^{2}+\Delta^{2}$.
Hence, the two-chain model (\ref{eq:H}) has a \emph{pair} of zero
modes at the edge, one for each value of $s_{z}$. These zero modes
form a single Kramers' pair. We have verified this explicitly (see Fig. \ref{fig:spectrum_B=0_TwoWires}).

\emph{Non-centrosymmetric perturbations.}$-$If inversion symmetry of the
quantum wire is broken, one expects the TSC phase to be robust over a finite range of
parameters as long as the system remains TRI.
As an example for such a perturbation we consider Rashba type spin orbit
coupling in a direction perpendicular to the plane of the wire, given by $\delta H=\lambda_{R}s_{y}\tau_{z}\sin k$.
It turns out that when $\lambda_{R}$ reaches a critical value of
the order of $\Delta$ the superconducting gap closes. As we show\cite{SM} this is due to the fact that
the pairing potential does not couple time-reversed states in this
case, but states related by inversion symmetry. Therefore a critical value
of the pairing potential is required to open a gap at the Fermi energy
once inversion symmetry is broken.

We conclude that in order to obtain the TSC phase, it is necessary
that the crystalline structure of the wire material is centrosymmetric.
In addition, the setup has to have an approximate inversion center,
such that Rashba-type spin orbit coupling is small.

Another question we address is what happens if the magnitude of the pairing potential on the two sides of the
wire is not exactly equal. The SC pairing can then be decomposed into odd ($\Delta_o$) and
even ($\Delta_e$) spatial components, such that the pairing gaps on the upper and lower chains are $\Delta_{1,2}=\Delta_e \pm \Delta_o$,
respectively. Once $\Delta_e$ is non-zero, the condition for TSC formulated
above is no longer satisfied. However, for small enough $\Delta_e$ we
expect the gap to remain finite and therefore the system remains in the topological phase.
Using the lattice Hamiltonian (\ref{eq:H}) and computing the phase diagram explicitly\cite{SM}
we find that as we increase $\Delta_e$ the gap remains open up to values of the order of a half of $\Delta_o$.
We thus conclude that our setup does not rely on the strength of the
proximity coupling to the two superconductors being equal.

\emph{Effect of TR breaking.}$-$Once time-reversal symmetry is broken
the two Majorana modes on each edge are no longer protected, and we
expect them to split from zero energy. In the suggested setup, TRI
can be broken either by changing the phase difference $\phi$ away
from $\pi$, or by applying a magnetic (Zeeman) field $\vec{B}$ in
the quantum wire. The Zeeman field is modelled by adding a term $-\vec{B}\cdot\vec{s}$ to the Hamiltonian $H_k$. The low energy effective Hamiltonian on the edge
can be written in terms of the local Majorana operators $\gamma_{1},\ \gamma_{2}$.
Denoting by $\lambda(\vec{B},\phi)$ the coupling between the two
Majoranas due to broken TRI, and demanding the Hamiltonian to be Hermitian
we conclude the coupling term must be of the form $\lambda i\gamma_{1}\gamma_{2}$.
Note that under time reversal $Ti\gamma_{1}\gamma_{2}T^{-1}=-i\gamma_{1}\gamma_{2}$.
Expanding $\lambda$ around $\vec{B}=0,\ \phi=\pi$ to lowest order
in both parameters, we see that the coupling must be of the form $\lambda(\vec{B},\phi)\propto\vec{B}\cdot\hat{n}+\alpha(\phi-\pi)$,
where $\hat{n}$ is some unit vector (note that all even orders in
the expansion must vanish due to TRI). This suggests that only a single
component of the magnetic field (parallel to $\hat{n}$) couples between
the Majorana modes. Moreover, for a given magnetic field we can vary
the flux and bring the coupling back to zero.

Using the lattice model (\ref{eq:H}) and calculating numerically
the BdG spectrum of a finite system, we have confirmed these results.
We find that to linear order in the magnetic field, only the $z$ component of the field leads to shifting of the edge energy levels away from zero at $\phi=\pi$. For non-zero
$B_{z}$ we vary the relative phase $\phi$ between the SC pairings
on the two opposite sides of the nanowire and plot the energy spectrum
obtained (see Fig. \ref{fig:spectrum_B!=0_TwoWires}).
It is clearly seen that when $B_{z}\ne0$ the zero crossings are shifted
away from $\phi=\pi$.

\emph{Adiabatic pumping.}$-$One can now consider an adiabatic cycle
in parameter space where the phase $\phi$ changes by $2\pi$. From
the arguments above, we expect a single level crossing to occur at
each edge for some value of $\phi$ which depends on $B_{z}$\cite{comment-Bz}.
The two states that cross differ by their local fermion parity, and therefore
they cannot mix. We argue that such a cycle falls into the non-trivial
class of adiabatic cycles in 1D particle-hole symmetric Hamiltonians
(class D), discussed in Refs. \cite{roy2011topological,teo2010topological},
and serves as a \emph{fermion parity pump}. Indeed,
one can define the parity of the right (left) edge as $P_{R,L}=\underset{i}{\prod}\left(-1\right)^{n_{i}}$,
where $n_{i}=c_{i}^{\dagger}c_{i}$ is the occupation of site $i$,
letting $i$ run over all the sites in the right (left) half of the
wire. An adiabatic sweep through the cycle changes the fermion parity
at each edge, i.e. takes the expectation value $\left\langle P_{R,L}\right\rangle $
to $-\left\langle P_{R,L}\right\rangle $ respectively.
One can construct an explicit $\mathbb{Z}_{2}$ topological invariant characterizing the
pumping process and show that it is non-zero for the cycle considered\cite{SM}.

In situations in which one component of the total spin is
conserved, e.g. $S_{z}$, the cycle also
pumps a quantum of spin angular momentum $S_{z}=1/2$ between the
two edges. Repeating this cycle twice is equivalent to the $\mathbb{Z}_{2}$
spin pump discussed in Ref. \cite{Shindou2005,Fu2006}. The spin pumping
property can be used as an experimental signature of the anomalous
edge states. At $\phi=\pi$ each edge supports two degenerate (many-body) states
with an opposite expectation value of $S_{z}$.
Since the two states differ by adding a single electron or hole, they
must have $\langle S_{z}\rangle=\pm1/4$ \cite{SM}.
When $\phi$ is changed adiabatically by $2\pi$, the local
spin of the edge switches. If $S_{z}$ is not conserved, the unit
of spin transferred between the edges during the adiabatic cycle is
not quantized; however, we still expect $\langle S_{z}\rangle$ of
each edge to flip its sign over one cycle.

\begin{figure}
\subfloat[\label{fig:QSH}]{\includegraphics[scale=0.5]{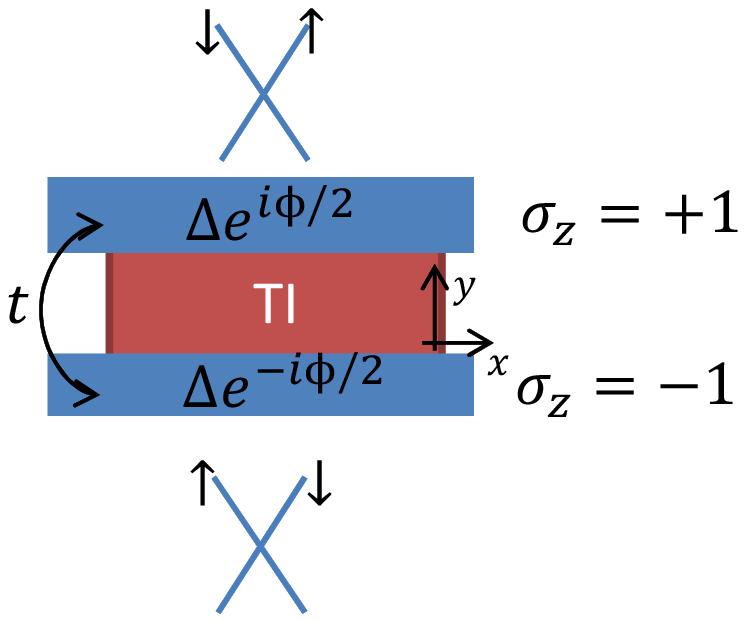}}
\subfloat[\label{fig:path}]{\includegraphics[scale=0.8]{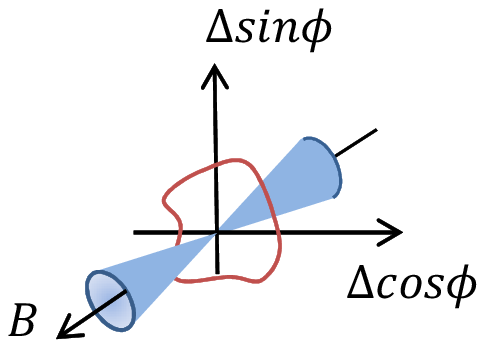}}

\caption{(a) 2D QSH model with finite tunneling probability between the edges.
(b) A closed path in parameters space encircling the gapless region
(cones) corresponding to a fermion parity pump. The path is not contractible
due to the persistence of the gapless region for $B\neq0$.}
\end{figure}

The pumping property becomes particularly transparent if one considers
an alternative model, illustrated in Fig. \ref{fig:QSH}. Consider
a strip of a 2D quantum spin Hall (QSH) material with 1D helical edge states.
If the width of the strip is finite, the tunneling amplitude $t_{\perp}$
between the edge states is non-zero. The opposite sides of the strip are proximity-coupled to two s-wave SCs with a phase difference of $\phi$.

In absence of a magnetic field and in the $t_{\perp}\rightarrow0$
limit, a cycle in which $\phi$ changes by $2\pi$ can be realized by passing
a superconducting vortex through the QSH strip (between the two SCs),
along the $x$ direction. Such a vortex induces a voltage along
the $y$ direction, which in its turn will lead to a spin current
along the $x$ direction. The total spin transferred between the ends of the QSH strip in this process is $1/2$, corresponding to a single fermion. Hence, such a cycle exactly serves as a fermion parity pump.
Note that the use of a QSH is not essential for the pumping phenomena.
In the QSH model, however, the origin of the pumping is evident.

Denoting the two edges of the QSH state by $\sigma_{z}=\pm1$, we
can write the following low energy effective Hamiltonian:
\begin{equation}
H=\left(vks_{z}\sigma_{z}-t_{\perp}\sigma_{x}-\mu\right)\tau_{z}+Bs_{z}+\Delta\cos\frac{\phi}{2}\tau_{x}+\Delta\sin\frac{\phi}{2}\sigma_{z}\tau_{y}.
\end{equation}
Here, $v$ is the velocity of the edge modes, $\mu$ is their chemical
potential, $B$ is an applied Zeeman field, and $\Delta$ is the induced
pairing potential. We examine the phase diagram of the system in the
parameter space spanned by $\Delta$, $\phi$ and $B$, Fig. \ref{fig:path}.
For $B=0$ and $\phi=0,\pi$, the system
is TRI. For $\mu>t_\perp$, the gapless point $\Delta=0$ separates between
the trivial and the topological phases. When a magnetic field is turned
on, the gapless point does not disappear but turns into a finite region
$|\Delta|\leq\left|B\right|$. As we change $\phi$ by $2\pi$,
the path in parameter space encircles a gapless region and can not be contracted to a
point without crossing it. This is a consequence of the fermion parity pumping property of this cycle\cite{comment-cycle}.

\emph{Discussion}.$-$We have presented a general setup to realize
a time reversal invariant TSC by proximity coupling a quantum wire
with strong SOC to conventional superconductors. The TSC phase can
be identified by the presence of a pair of zero-energy Majorana bound
states at each edge, protected by time-reversal symmetry. Thus, we
expect a zero-bias peak to appear in the tunneling conductance into
the edge of the system when the phase difference between the two superconductors
is $\phi=\pi$. Intriguingly, varying $\phi$ adiabatically by $2\pi$
pumps both fermion parity and spin between the edges.

\emph{Acknowledgements.}$-$We thank A. Akhmerov for useful discussions.
E. B. was supported by the Israel Science Foundation under Grant 7113640101, and by the Robert Rees Fund.
A. S. thanks the US-Israel Binational Science Foundation, the Minerva
foundation, and Microsoft Station Q for financial support.

\bibliography{ref}

\begin{thebibliography}{42}
\bibitem[{com({\natexlab{a}})}]{comment-Bz}
\bibinfo{note}{From the form $\lambda\propto \vec{B}\cdot \hat{n}+\alpha
  (\phi-\pi)$, one can see that this statement holds for sufficiently small
  $|\vec{B}|$. In fact, it is true for an arbitrary magnetic field as long as
  the system remains gapped in the bulk. This follows from the pumping
  properties of the cycle.}
\end{thebibliography}

\begin{widetext}
\section*{Supplementary Material}

\section{\emph{Obtaining a TRI TSC phase}}

\subsection{Conditions for a centrosymmetric system to be in the topological
phase\label{sec:Conditions-for-TRI}}

We show that a centrosymmetric time-reversal invariant 1D system is
necessarily in the topological phase if it satisfies the conditions
stated in the main text, i.e. $(1)$ odd-parity pairing with full
superconducting gap and $\left(2\right)$ the number of (spin-degenerate)
pairs of Fermi points between $k=0$ and $k=\pi$ in the normal state
is odd. This was proved in Ref. \cite{Fu2010} for 2D and 3D systems.

A Fermi surface invariant for one-dimensional TRI topological superconductors
was derived by Qi et. al. \cite{Qi2010}, assuming that
each Fermi point is singly degenerate. The invariant is then equal
to the product of the signs of the pairing potential on each Fermi
point. However, in TRI systems with centrosymmetry each Fermi point
is at least doubly degenerate. This is due to the fact that the operator
$PT$ (time-reversal followed by parity) commutes with the normal
state Hamiltonian $h_{k}$ and that $\left(PT\right)^{2}=-1$. Hence,
by Kramers' theorem, for each eigenstate $\left|\Psi_{k_{F},1}\right\rangle $
of $h_{k}$, $\left|\Psi_{k_{F},2}\right\rangle =PT\left|\Psi_{k_{F},1}\right\rangle $
is an orthogonal eigenstate with the same energy. We therefore first
generalize the formulation of the invariant given in \cite{Qi2010}
for the case of degenerate Fermi points and then use it to prove the
criterion stated above. 

Recall that in the presence of particle-hole and time-reversal symmetry,
a BdG Hamiltonian $H=\left(\begin{array}{cc}
h_{k} & \Delta_{k}\\
\Delta_{k}^{\dagger} & -h_{-k}^{*}
\end{array}\right)$ can be brought to the form $H=\left(\begin{array}{cc}
0 & Q_{k}\\
Q_{k}^{\dagger} & 0
\end{array}\right)$ where $Q_{k}=h_{k}+i\tilde{T}\Delta_{k}$ ($\tilde{T}=is_{y}$ is
the unitary part of the time-reversal operator) \cite{Schnyder2008}. 

Denote the eigenstates of $h_{k}$ by $\left|n,k,\alpha\right\rangle $
where $n$ denotes the energy bands and $\alpha$ distinguishes between
different degenerate eigenstates. In the weak pairing limit $\Delta$
couples only degenerate eigenstates of $h_{k}$, and $Q_{k}$ can
be approximated as 
\begin{equation}
Q_{k}\simeq\underset{n,k,\alpha,\alpha'}{\sum}\left[\epsilon_{nk}\delta_{\alpha,\alpha'}+i\left(\tilde{T}\Delta_{k}\right)_{\alpha,\alpha'}\right]\left|\Psi_{n,k,\alpha}\right\rangle \left\langle \Psi_{n,k,\alpha'}\right|,
\end{equation}
where $\left(\tilde{T}\Delta_{k}\right)_{\alpha,\alpha'}=\left\langle \Psi_{n,k,\alpha}\right|\tilde{T}\Delta_{k}\left|\Psi_{n,k,\alpha'}\right\rangle $
is a Hermitian matrix. 

%\begin{comment}
%Due to TRS $T\Delta_{k}T^{-1}=\Delta_{-k}$. Using also $\Delta_{k}^{T}=-\Delta_{-k}\Rightarrow\Delta_{k}^{\star}=-\Delta_{-k}^{\dagger}$
%we can show that $\Delta_{k}s_{y}$ is anti-hermitian:
%
%$\left(is_{y}K\right)\Delta_{k}\left(-is_{y}K\right)=s_{y}\Delta_{k}^{*}s_{y}=\Delta_{-k}\Rightarrow s_{y}\Delta_{k}s_{y}=\Delta_{-k}^{\star}=-\Delta_{k}^{\dagger}\Rightarrow\Delta_{k}s_{y}=-s_{y}\Delta_{k}^{\dagger}=-\left(\Delta_{k}s_{y}\right)^{\dagger}$.
%
%Therefore $\Delta_{k}is_{y}$ is Hermitian.
%\end{comment}

One can diagonalize it and write 
\begin{equation}
Q_{k}\simeq\underset{n,k,\beta}{\sum}\left[\epsilon_{nk}+i\delta_{nk\beta}\right]\left|\psi_{n,k,\beta}\right\rangle \left\langle \psi_{n,k,\beta}\right|,
\end{equation}
where $\delta_{n,k,\beta}$ are the eigenvalues of the matrix $\tilde{T}\Delta_{k}$,
and $\left|\psi_{n,k,\beta}\right\rangle $ are its eigenstates. The
topological invariant for 1D TRI TSC is now $\underset{j,\beta}{\prod}sgn\left(\delta_{n_{j},k_{F,j},\beta}\right)$
where the product runs over all the Fermi points between $0$ and
$\pi$ (labeled by $j$), and $n_{j}$, $k_{F,j}$ are the band index
and Fermi momentum of the $j$th Fermi point, respectively. 

We now prove that for odd-parity pairing, i.e. $P\Delta_{k}P=-\Delta_{-k}$,
if the chemical potential crosses an odd number of (doubly degenerate)
bands, the topological invariant is $-1$ and the system is in the
topological phase.

As was already mentioned, in a centrosymmetric system the eigenstates
of $h_{k}$ come in pairs: $\left|\Psi_{k_{F},1}\right\rangle $,
$\left|\Psi_{k_{F},2}\right\rangle =PT\left|\Psi_{k_{F},1}\right\rangle $.
Therefore, the dimension of the matrix $\tilde{T}\Delta_{k}$ at each
Fermi point is even. Note also that 
\begin{equation}
\left(\tilde{T}\Delta_{k}\right)\left(PT\right)=-\left(PT\right)\left(\tilde{T}\Delta_{k}\right).
\end{equation}
This means that $\left\langle \Psi_{k,1}\right|\tilde{T}\Delta_{k}\left|\Psi_{k,1}\right\rangle =-\left\langle \Psi_{k,2}\right|\tilde{T}\Delta_{k}\left|\Psi_{k,2}\right\rangle $,
i.e. the matrix $\tilde{T}\Delta_{k}$ is traceless. Therefore, the
eigenvalues $\delta_{n,k,\beta}$ come in pairs, $\delta_{n,k,\beta}=\pm\lambda$,
and for an odd number of pairs of Fermi points between $0$ and $\pi$,
the product $\underset{j,\alpha}{\prod}sgn\left(\delta_{n_{j},k_{F,j},\beta}\right)$
is equal to $-1$.

\newpage{}

\subsection{Spin-orbit coupling}

Spin-orbit coupling is not mentioned explicitly in the conditions
discussed in section \ref{sec:Conditions-for-TRI} above. However,
we argue that it is an essential ingredient for the realization of
a TRI TSC phase. Indeed, consider a system with no spin-orbit interaction.
As the system is also TRI, we must have full spin rotational symmetry.
It is enough however to assume that only two spin components, e.g.
$s_{y}$ and $s_{z}$, are conserved. A particle-hole symmetric system
satisfies $CHC^{-1}=-H$ where $H$ is the BdG Hamiltonian in Nambu
spinor basis and $C=s_{y}\tau_{y}K$ is the particle-hole transformation
operator (here $K$ denotes complex conjugation). For spin singlet
SC pairing, conservation of $s_{y}$ allows one to define an operator
$\tilde{C}=s_{y}C=\tau_{y}K$ which also anti-commutes with the Hamiltonian
but satisfies $\tilde{C}^{2}=-1$. Assume the system has a zero-energy
edge state $H\left|\Psi\right\rangle =0$. Since $\left[H,s_{z}\right]=0$
one can choose $\left|\Psi\right\rangle $ to be an eigenstate of
$s_{z}$, say with spin up. The state $\tilde{C}\left|\Psi\right\rangle $
is also a zero-energy eigenstate, orthogonal to $\left|\Psi\right\rangle $.
Moreover, $\tilde{C}$ is an operator that is local and does not flip
the spin of the state. Hence, we obtained two zero-energy states on
the same edge with the same spin. There is no symmetry that protects
the states from splitting and they will generically be shifted away
from zero energy. This proves that a spin conserving system must be
in the trivial phase.

\subsection{Breaking inversion symmetry}

\subsubsection{Non-cenotrsymmetric nanowire}

In the main text we discuss what happens if a perturbation that breaks
inversion symmetry of the non-superconducting part of the Hamiltonian
is present. Since parity symmetry is not necessary for the appearance
of the TRI TSC phase, we expect the system to remain in the topological
state as long as the perturbation is small and the system remains
gapped. We then ask how big does the perturbation have to be to drive
the system out of the topological phase. As an example for such a
perturbation we consider Rashba type spin orbit coupling in a direction
perpendicular to the plane of the wire given by $\delta H=\lambda_{R}s_{y}\tau_{z}\sin k$.
It turns out that the superconducting gap closes when $\lambda_{R}$
reaches a critical value of the order of $\Delta$. This happens due
to the fact that the pairing potential cannot couple time-reversed
states in this case but only states related by inversion as we show
below. Therefore a critical value of the pairing potential is required
to open a gap at the Fermi energy once inversion symmetry is broken. 

To see which states can the SC pairing couple, note that the system
possesses mirror symmetry w.r.t. the $y$ axis: $M_{y}=s_{y}\sigma_{x}$,
i.e. $\left[H_{0k},M_{y}\right]=0$. The SC pairing $\Delta$ is odd
under this symmetry, i.e. $\left\{ H_{\Delta},M_{y}\right\} =0$.
Consider an eigenstate $\left|\Psi_{+}\right\rangle $ of $H_{0k}$
at the Fermi momentum with an $M_{y}$ eigenvalue of $+1$. When parity
symmetry is broken, there is generically no additional degenerate
state at the same energy and momentum. The SC pairing can only couple
between a particle state $\left|\Psi_{+}\right\rangle $ and its time-reversed
hole state with an opposite momentum, whose wavefunction is $\tau_{y}\left|\Psi_{+}\right\rangle $.
Note that the mirror eigenvalue of $\tau_{y}\left|\Psi_{+}\right\rangle $
is the same as that of $\left|\Psi_{+}\right\rangle $. As a result,
the matrix element corresponding to the SC pairing part of $H_{k}$,
i.e. $H_{\Delta}\tau_{x}$, between these states vanishes: $\left\langle \Psi_{+}\right|\tau_{y}H_{\Delta}\tau_{x}\left|\Psi_{+}\right\rangle =\left\langle \Psi_{+}\right|\tau_{y}M_{y}H_{\Delta}\tau_{x}M_{y}\left|\Psi_{+}\right\rangle =-\left\langle \Psi_{+}\right|\tau_{y}H_{\Delta}\tau_{x}\left|\Psi_{+}\right\rangle $.
If parity symmetry is present, however, there is another degenerate
eigenstate at the Fermi momentum given by $\vert\Psi_{-}\rangle=PT\vert\Psi_{+}\rangle=\sigma_{x}is_{y}K\vert\Psi_{+}\rangle$.
Note that the mirror eigenvalues of the two degenerate states are
\emph{opposite}. The superconducting pairing between different $M_{y}$
eigenvalue states $\left\langle \Psi_{-}\right|\tau_{y}H_{\Delta}\tau_{x}\left|\Psi_{+}\right\rangle $
is generically non-zero and a superconducting gap opens even for small
$\Delta$.

\subsubsection{Unequal superconducting pairing}

Another question we address is what happens if the pairing potential
has a component which is even under inversion. The pairing part of
the Hamiltonian is then replaced by $H'_{\Delta}=\Delta_{\mathrm{e}}+\Delta_{\mathrm{o}}\sigma_{z}$,
where $\Delta_{e}$ and $\Delta_{o}$ are the even and odd components
respectively. Once $\Delta_{\mathrm{e}}$ is non-zero, the condition
for TSC formulated above is no longer satisfied. However, for small
enough $\Delta_{\mathrm{e}}$ we expect the gap to remain finite and
therefore the system remains in the topological phase. The phase diagram
obtained by varying $\mu$ and $\Delta_{\mathrm{e}}$ for fixed $\Delta_{\mathrm{o}}$
is shown in Fig. \ref{fig:gap}. For $\Delta_{\mathrm{e}}=0$ and
$\mu$ below the lowest (doubly degenerate) energy band we start in
the trivial phase, but as we increase $\mu$ and cross the bottom
of the lowest band the gap closes and re-opens. In this region, we
observe the zero-energy edge states mentioned earlier, and therefore
the system is in the topological phase. As we increase $\Delta_{e}$
the gap remains open, up to values of the order of a half of $\Delta_{\mathrm{o}}$.
In this entire region the system remains in the topological phase. 

\begin{figure}
\noindent \subfloat[]{\includegraphics[scale=0.2]{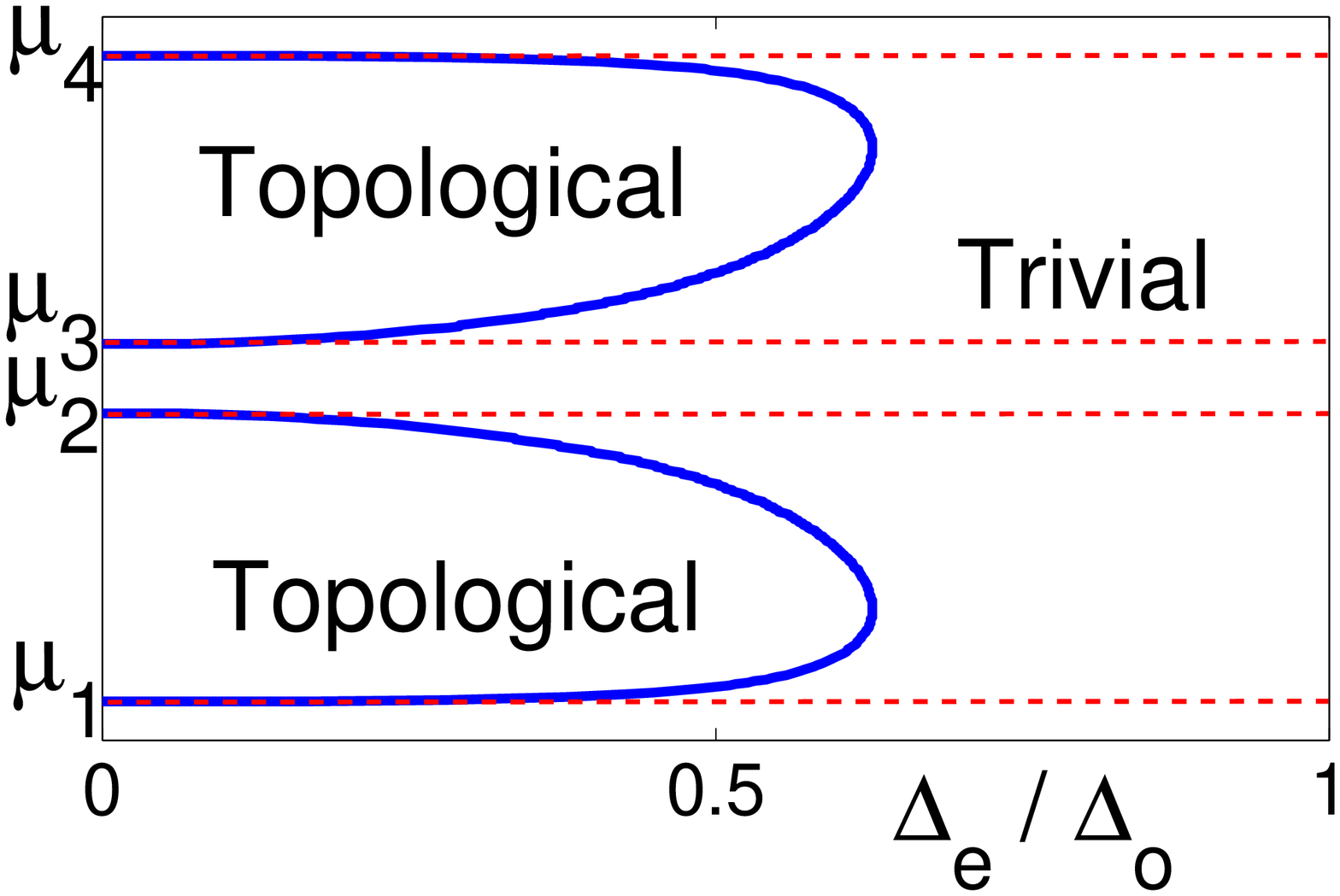}

}\subfloat[]{\includegraphics[scale=0.2]{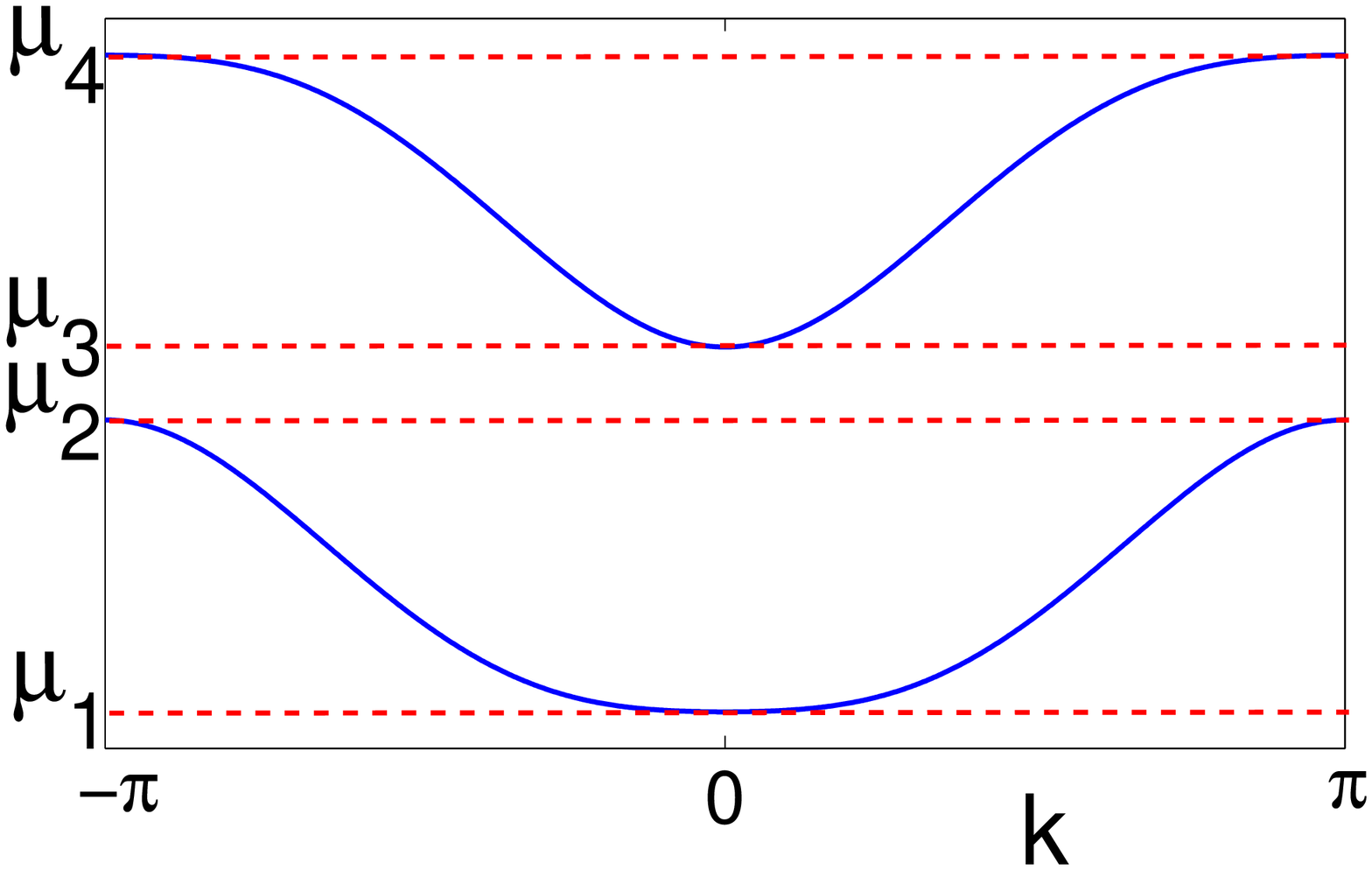}

}

\caption{(a) Phase diagram of the system letting the chemical potential $\mu$
and the ratio between even and odd pairing components $\Delta_{e}/\Delta_{o}$
vary. Along the blue line the superconducting gap closes and the system
undergoes a transition between the topological and the trivial phases.
(b) Energy bands of the system. For $\Delta_{e}=0$ a phase transition
to the topological phase occurs as the chemical potential crosses
the bottom of each band. For both plots, the nanowire is taken to
be infinite in the $x$ direction with lattice model parameters $t_{\perp}=2.5,\ t=1,\ \lambda=1$,
and odd pairing $\Delta_{o}=0.4$.\label{fig:gap}}
\end{figure}

\section{\emph{Expectation value of the spin on the edge of a system in the
TSC phase}}

We would like to show that a system in the topological TRI phase can
have a non-zero spin expectation value at the edges. This is most
clearly seen for a spin conserving system (we will assume that the
$z$ component of the spin is conserved). In this case, the expectation
value of the spin at each edge is $1/2$ of an electron spin, $\langle S_{z}\rangle=\pm1/4$.

To see this recall that the edge of a system hosts two Majorana zero
modes $\gamma_{1},\gamma_{2}$. These two Majoranas form a single
Fermionic degree of freedom $\Psi=\gamma_{1}+i\gamma_{2}$. Creation,
or annihilation, of this Fermionic excitation transfers the system
between the two degenerate ground states of the edge. Since $S_{z}$
is a good quantum number, we can assign a well defined spin to this
Fermionic mode, say $\uparrow$, and denote it $\Psi_{\uparrow}$.
We therefore see that the expectation value of the spin in the two
ground states $\left|GS_{1}\right\rangle $ and $\left|GS_{2}\right\rangle =\Psi_{\uparrow}^{\dagger}\left|GS_{1}\right\rangle $
must differ by spin of a single fermion, $\langle S_{z}\rangle_{GS_{2}}-\langle S_{z}\rangle_{GS_{1}}=\frac{1}{2}$.
Furthermore, the two ground states are related by time-reversal symmetry,
and should therefore have opposite spin expectation values $\langle S_{z}\rangle_{GS_{2}}=-\langle S_{z}\rangle_{GS_{1}}$.
Combining the two, we get that at the edge, $\langle S_{z}\rangle=\pm1/4$
for the two ground states.

The non-zero spin expectation value at the edge of a system is a direct
consequence of the time-reversal anomaly discussed by Chung et al.
\cite{chung2012time}. For a system in the TRI topological phase,
time-reversal operation changes the local fermion parity at its edge
(note that the parity of the whole system remains unchanged). However,
while the parity of an edge is not an easy observable to measure,
the expectation value of the local spin density on the edge suggests
a more practical way to observe the anomaly.

\section{\emph{Parity pump $\mathbb{Z}_{2}$ invariant}}

In the main text we argued that a process in which the relative phase
between the two superconductors on the two sides of the 1D wire is
varied $\phi:0\rightarrow2\pi$, falls into the non-trivial class
of adiabatic cycles in 1D particle-hole symmetric Hamiltonians (class
D). We now construct an explicit bulk $\mathbb{Z}_{2}$ topological
invariant characterizing the process and show that it is non-zero
for the cycle considered. For this purpose we use the model of a centrosymmetric
wire with spin orbit coupling considered in the main text. It should
be noted that the constructed invariant requires a continuous single
valued gauge choice for the eigenstates on the half torus $k\in\left[0,\pi\right],\ \phi\in\left[0,2\pi\right]$. 

We first formulate the invariant and show that it is unchanged by
a unitary transformation that mixes between the negative energy bands.
We then give an explicit way of calculating the invariant and demonstrate
it on the system considered in the main text.

\subsection{Formulation of the invariant}

Given a many-body Hamiltonian $\mathcal{H}$, particle-hole symmetry
implies the existence of an anti-unitary operator $\mathcal{C}$,
such that $\mathcal{C}\mathcal{H}\mathcal{C}^{-1}=-\mathcal{H}$.
For a non-interacting periodic system the Hamiltonian is given by
$\mathcal{H}=\sum\Psi_{k}^{\dagger}H\left(k\right)\Psi_{k}$ . Particle-hole
symmetry can then be expressed in terms of the Bloch Hamiltonian $H\left(k\right)$
as $CH\left(-k\right)C^{-1}=-H\left(k\right)$, where $C$ is an antiunitary
operator acting in the single particle basis. Note that it can be
written as a composition of a unitary operator and the complex conjugation
operation that we denote by $K$.

For a BdG Hamiltonian in the Nambu spinor basis $\Psi_{k}^{\dagger}=\left(c_{k}^{\dagger},-is_{y}c_{-k}\right)$
one can show that the particle-hole operator is given by $C=s_{y}\tau_{y}K$,
where $s_{y}$ and $\tau_{y}$ are Pauli matrices in the spin and
particle-hole subspaces respectively. A unitary transformation $U=e^{i\frac{\pi}{4}\left(1-s_{y}\tau_{y}\right)}$
transforms $C$ to $\tilde{C}=UCU^{\dagger}=K$ and $H\left(k\right)$
to $\tilde{H}\left(k\right)=UH\left(k\right)U^{\dagger}$. In the
new basis, particle-hole symmetry implies that $\tilde{C}\tilde{H}\left(-k\right)\tilde{C}^{-1}=\tilde{H}^{*}\left(-k\right)=-\tilde{H}\left(k\right)$.
At the TRI momenta $k=0,\pi$ we have $\tilde{H}_{0,\pi}=-\tilde{H}_{0,\pi}^{*}$,
meaning that $\tilde{H}$ can be written as $\tilde{H}_{0,\pi}=iA_{0,\pi}$
where $A_{0,\pi}$ are real anti-symmetric matrices. Note that at
$k=0,\pi$ the transformation $U$ is simply a transformation to the
Majorana basis.

The dimension of the BdG Hamiltonian (and hence also of $A$) is even
and we denote it by $2n$. Any real anti-symmetric matrix of even
dimension can be brought to the canonical, block-diagonal, form 
\[
\tilde{A}=\left(\begin{array}{cccccc}
0 & E_{1}\\
-E_{1} & 0\\
 &  & ...\\
 &  &  & ...\\
 &  &  &  & 0 & E_{n}\\
 &  &  &  & -E_{n} & 0
\end{array}\right)
\]
 by an orthogonal transformation $V\in O\left(2n\right)$, where $\left\{ \pm E_{i}\right\} _{i=1..n}$
is the set of eigenenergies of $H$, with $\left\{ E_{i}\right\} _{i=1..n}>0$.

Denoting the columns of $V_{0,\pi}$ by $\left\{ u_{i}^{0,\pi},v_{i}^{0,\pi}\right\} _{i=1..n}$
the negative energy eigenstates of $H_{0,\pi}$ are given by $\left\{ \Psi_{i}^{0,\pi}=u_{i}^{0,\pi}-iv_{i}^{0,\pi}\right\} _{i=1..n}$.

Now consider an adiabatic cycle parametrized by $\phi:0\rightarrow2\pi$
(in the cycle considered in the main text $\phi$ is the relative
phase between the superconductors). The BdG Hamiltonian at each $k$
varies with $\phi$. At $k=0,\pi$ the Hamiltonian can still be written
as $\tilde{H}_{0,\pi}\left(\phi\right)=iA_{0,\pi}\left(\phi\right)$
where $A_{0,\pi}\left(\phi\right)$ are real matrices for all $\phi$
and can be brought to the canonical form mentioned above by $V_{0,\pi}\left(\phi\right)\in O\left(2n\right)$.
We assume that $V_{0,\pi}\left(\phi\right)$ can be chosen to be continuous
in $\phi$ (this is equivalent to requiring a continuous gauge choice
for the eigenstates $\Psi_{i}^{0,\pi}\left(\phi\right)$).

Denoting $W_{0,\pi}\left(\phi\right)=V_{0,\pi}\left(\phi=0\right)^{-1}V_{0,\pi}\left(\phi\right)\in SO\left(2n\right)$
we find that the adiabatic cycle $\phi:0\rightarrow2\pi$ corresponds
to a closed path in $SO\left(2n\right)$. Since the homotopy group
$\pi_{1}\left(SO\left(N\right)\right)$, for $N\geq3$, is isomorphic
to $\mathbb{Z}_{2}$ we can assign to $W_{0,\pi}\left(\phi\right)$
a $\mathbb{Z}_{2}$ index, $\nu$. We argue that the difference between
the indices corresponding to the paths $W_{0}\left(\phi\right)$ and
$W_{\pi}\left(\phi\right)$ , $\Delta\nu=\nu_{\pi}-\nu_{0}$, is the
invariant characterizing the parity pumping process.

\subsection{Unitary transformations in the negative energy subspace}

Since all topological invariants should be independent of the spectrum
of the Hamiltonian as long as it remains gapped, we can flatten the
energy bands (such that all $E_{i}=1$) as long as we keep the eigenstates
unmodified. We then expect $\Delta\nu$ to be invariant under a unitary
transformation in the subspace of negative energy states $\left\{ \Psi_{i}\left(k,\phi\right)\right\} _{i=1..n}$.
We now show that this is indeed the case.

Consider a transformation $\tilde{\Psi}\left(k,\phi\right)=R\left(k,\phi\right)\Psi\left(k,\phi\right)$
where $\Psi\left(k,\phi\right)$ is an $n$-dimensional vector and
$R\left(k,\phi\right)\in U\left(n\right)$ is continuous for all $k\in\left[0,\pi\right],\ \phi\in\left[0,2\pi\right]$.
At $k=0,\pi$ a transformation of $\Psi$ corresponds to a transformation
of $V$ (recall that the columns of $V$ are simply the real and imaginary
parts of the eigenstates $\left\{ u_{i},v_{i}\right\} _{i=1..n}$).
To find an explicit form for this transformation consider $V_{ord}\left(\phi\right)=B\left(\phi\right)V\left(\phi\right)B^{T}\left(\phi\right)$
where $B\left(\phi\right)$ is an orthogonal transformation that changes
the order of columns of $V\left(\phi\right)$, such that $V_{ord}\left(\phi\right)=\left(\begin{array}{cc}
| & |\\
u_{i=1..n} & v_{i=1..n}\\
| & |
\end{array}\right)$. As we argue in section \ref{sub:Lemma} below, such a transformation
does not change the $\mathbb{Z}_{2}$ index of the closed path. Now
decompose $R\left(k,\phi\right)$ into its real and imaginary parts
$R\left(k,\phi\right)=R_{1}\left(k,\phi\right)+iR_{2}\left(k,\phi\right)$.
Then the transformation at $k=0,\pi$ can be written as $\tilde{V}_{ord}=V_{ord}M$
with $M=\left(\begin{array}{cc}
R_{1} & R_{2}\\
-R_{2} & R_{1}
\end{array}\right)$. Note that even though $V$ and $V_{ord}$ are defined only for $k=0,\pi$,
the matrix $M$ is well defined and continuous for all $k\in\left[0,\pi\right],\ \phi\in\left[0,2\pi\right]$.
One can also check that unitarity of $R\left(k,\phi\right)$ leads
to orthogonality of $M\left(k,\phi\right)$.

%\begin{comment}
%Consider a $2n\times n$ matrix formed by the $n$ occupied eigenstates
%of the $2n$ dimensional BdG Hamiltonian,$\Psi_{ji}\left(k,\phi\right)$
%, $j=1..2n$ and $i=1..n$.
%
%A change of basis can mix between these bands $\tilde{\Psi}_{ki}\left(k,\phi\right)=\Psi_{ij}\left(k,\phi\right)U_{jk}\left(k,\phi\right)$
%with $U\left(k,\phi\right)\in U\left(n\right)$. Decomposing $U$
%into its real and imaginary parts $U=A+iB$ we obtain for $O\left(\phi\right)$:
%\begin{equation}
%\tilde{O}\left(\phi\right)=\left(\begin{array}{cc}
%| & |\\
%\tilde{u}_{i} & \tilde{v}_{i}\\
%| & |
%\end{array}\right)=\left(\begin{array}{cc}
%| & |\\
%u_{i} & v_{i}\\
%| & |
%\end{array}\right)\left(\begin{array}{cc}
%A & B\\
%-B & A
%\end{array}\right)\equiv O\left(\phi\right)K\left(\phi\right).
%\end{equation}
%
%
%Note that $K^{T}\left(\phi\right)K\left(\phi\right)=\mathbb{I}$,
%i.e. $K\left(\phi\right)\in O\left(2n\right)$:
%
%$U^{\dagger}U=\left(A+iB\right)^{\dagger}\left(A+iB\right)=\left(A^{T}-iB^{T}\right)\left(A+iB\right)=A^{T}A+B^{T}B+i\left(A^{T}B-B^{T}A\right)=\mathbb{I}\ \Rightarrow\ A^{T}A+B^{T}B=\mathbb{I},\ A^{T}B-B^{T}A=0$
%
%\[
%K^{T}\left(\phi\right)K\left(\phi\right)=\left(\begin{array}{cc}
%A^{T} & -B^{T}\\
%B^{T} & A^{T}
%\end{array}\right)\left(\begin{array}{cc}
%A & B\\
%-B & A
%\end{array}\right)=\left(\begin{array}{cc}
%A^{T}A+B^{T}B & A^{T}B-B^{T}A\\
%B^{T}A-A^{T}B & B^{T}B+A^{T}A
%\end{array}\right)=\left(\begin{array}{cc}
%\mathbb{I} & 0\\
%0 & \mathbb{I}
%\end{array}\right)=\mathbb{I}_{2n\times2n}
%\]
%\end{comment}

Overall, we have

\begin{equation}
\tilde{V}\left(\phi\right)\sim\tilde{V}_{ord}\left(\phi\right)=V_{ord}\left(\phi\right)M\left(\phi\right)\sim V\left(\phi\right)M\left(\phi\right),
\end{equation}

where the equivalence relation denoted by $\sim$ means that the two
paths are homotopic and therefore belong to the same element of $\pi_{1}\left(SO\left(2n\right)\right)$
(have the same $\mathbb{Z}_{2}$ index).

\begin{equation}
\tilde{W}_{0,\pi}\left(\phi\right)=\tilde{V}_{0,\pi}\left(\phi=0\right)^{-1}\tilde{V}_{0,\pi}\left(\phi\right)\sim W_{0,\pi}\left(\phi\right)M_{0,\pi}\left(\phi\right).
\end{equation}

As we show in \ref{sub:Lemma}, given two paths $O_{1}\left(\phi\right),\ O_{2}\left(\phi\right)$
in $SO(N)$ with $\mathbb{Z}_{2}$ indices $\nu_{1},\ \nu_{2}$ respectively,
the path $O\left(\phi\right)=O_{1}\left(\phi\right)O_{2}\left(\phi\right)$
has $\mathbb{Z}_{2}$ index $\nu=\nu_{1}+\nu_{2}$. Applying this
to $\tilde{W}_{k=0,\pi}$ we find that its $\mathbb{Z}_{2}$ index
is $\tilde{\nu}_{0,\pi}=\nu_{0,\pi}+\nu_{M_{0,\pi}}$ for $k=0,\pi$
respectively, where $\nu_{K_{0,\pi}}$ is the index corresponding
to $M_{0,\pi}\left(\phi\right)$. As was already mentioned, $M\left(k,\phi\right)$
is continuous for all $k\in\left[0,\pi\right]$, which means that
the two paths $M_{0}\left(\phi\right)$ and $M_{\pi}\left(\phi\right)$
can be continuously deformed into each other. Therefore, their corresponding
$\mathbb{Z}_{2}$ index $\nu_{M}$ is the same for $k=0,\pi$. Hence,
even though the indices $\nu_{0,\pi}$ can change due to the transformation,
their difference remains invariant $\tilde{\nu}_{\pi}-\tilde{\nu}_{0}=\nu_{\pi}-\nu_{0}$.

It might be instructive to consider an example in which $\nu_{M}\neq0$
and as a result each of the indices $\nu_{0,\pi}$ is changed during
the transformation. Consider one of the negative energy eigenstates
$\Psi_{i}\left(\phi\right)=u_{i}\left(\phi\right)-iv_{i}\left(\phi\right)$.
The wave function is defined up to a phase, which we require to be
continuous in $\phi$. The transformation $\tilde{\Psi}_{i}=e^{i\phi}\left(u_{i}-iv_{i}\right)$,
in which the phase of the eigenstate winds once with $\phi$, is thus
a valid one. As will become clear in section \ref{sub:Explicit-way},
such a transformation corresponds to a path $M\left(\phi\right)$
with non-zero $\mathbb{Z}_{2}$ index.

\subsection{\label{sub:Explicit-way}Explicit way of calculating the invariant}

We show below that the $\mathbb{Z}_{2}$ index of a path given by
the $2n\times2n$ orthogonal matrix $W\left(\phi\right)$ can be obtained
from its eigenvalues. Recall that the eigenvalues of an orthogonal
matrix of an even dimension are of the form $e^{i\varphi}$ and come
in complex conjugate pairs. The invariant $\nu$ is given by the parity
of the winding number of $\varphi\left(\phi\right)\equiv\underset{i=1}{\overset{n}{\sum}}\varphi_{i}\left(\phi\right)$.
This is equivalent to plotting all the phases $\pm\varphi_{i}$ vs
$\phi$ and counting the number of crossings at $\varphi=\pi$. The
parity of the number of crossings gives the invariant $\nu$ of $W\left(\phi\right)$. 

To see this recall that any orthogonal matrix $W\in SO(2n)$ can be
brought, by an orthogonal transformation, to the canonical form which
we denote by $D$. The matrix $D$ is block diagonal, each block being
a $2\times2$ rotation matrix $d_{i}=\left(\begin{array}{cc}
\cos\varphi_{i} & \sin\varphi_{i}\\
-\sin\varphi_{i} & \cos\varphi_{i}
\end{array}\right)$ and $i$, the index of the block, runs from $1$ to $n$. Note, that
further transformation by a \emph{unitary} matrix is required to bring
each block to the diagonal form $\left(\begin{array}{cc}
e^{i\varphi_{i}} & 0\\
0 & e^{-i\varphi_{i}}
\end{array}\right)$.

Now consider a path $W\left(\phi\right)\in SO(2n)$. Assuming it can
be brought to the canonical form in a continuous way, $W\left(\phi\right)=P\left(\phi\right)D\left(\phi\right)P\left(\phi\right)^{T}$,
and using \ref{sub:Lemma}, we find that $W\left(\phi\right)$ is
homotopic to $D\left(\phi\right)$. It is therefore enough to calculate
the $\mathbb{Z}_{2}$ index of $D\left(\phi\right)$.

Denote by $D_{i}\left(\phi\right)$ the unity matrix with its $i$'th
$2\times2$ block replaced by $d_{i}\left(\phi\right)$. Then we can
write $D\left(\phi\right)$ as $D\left(\phi\right)=\underset{i}{\prod}D_{i}\left(\phi\right)$.
We would like to show that $D\left(\phi\right)$ can be continuously
deformed into $\tilde{D}_{1}\left(\phi\right)$ with $\tilde{\varphi}_{1}\left(\phi\right)=\underset{i}{\sum}\varphi_{i}\left(\phi\right)=\varphi\left(\phi\right)$.
This can be achieved by a sequence of continuous deformations: at
each step we gradually rotate the $\left(i+1\right)$'th plane of
rotation onto the $i$'th one. This corresponds to the transformation
$D_{i}\left(\phi\right)P\left(t\right)D_{i+1}\left(\phi\right)P\left(t\right)^{T}$
with $t\in\left[0,1\right]$ where $P\left(t=0\right)=\mathbb{I}$
and $P\left(t=1\right)$ is a rotation of the axis $x_{2i+1},\ x_{2i+2}$
onto the axis $x_{2i-1},\ x_{2i}$ respectively. At the end of the
deformation we can rewrite $D_{i}\left(\phi\right)P\left(t=1\right)D_{i+1}\left(\phi\right)P\left(t=1\right)^{T}=\tilde{D}_{i}\tilde{D}_{i+1}$
where now $\tilde{D}_{i+1}=\mathbb{I}$ and the $i$'th block of $\tilde{D}_{i}$
is $\tilde{d}_{i}=d_{i}d_{i+1}=\left(\begin{array}{cc}
\cos\tilde{\varphi}_{i}\left(\phi\right) & \sin\tilde{\varphi}_{i}\left(\phi\right)\\
-\sin\tilde{\varphi}_{i}\left(\phi\right) & \cos\tilde{\varphi}_{i}\left(\phi\right)
\end{array}\right)$ with $\tilde{\varphi}_{i}=\varphi_{i}+\varphi_{i+1}$.

It can be shown that the path $D\left(\phi\right)$ describes a non-contractible
path in $SO(2n)$ if and only if $D_{3}\left(\phi\right)=\left(\begin{array}{ccc}
\cos\varphi\left(\phi\right) & \sin\varphi\left(\phi\right) & 0\\
-\sin\varphi\left(\phi\right) & \cos\varphi\left(\phi\right) & 0\\
0 & 0 & 1
\end{array}\right)\in SO(3)$ describes a non-contractible path in $SO(3)$.%
%\begin{comment}
%http://math.stackexchange.com/questions/123650/fundamental-group-of-the-special-orthogonal-group-son
%\end{comment}
 In $SO\left(3\right)$ the latter path is not-contractible if $\varphi\left(\phi\right)$
performs an odd number of windings as $\phi$ is varied from $0$
to $2\pi$. We therefore deduce that the $\mathbb{Z}_{2}$ index is
given by the parity of the winding number of $\varphi\left(\phi\right)=\underset{i}{\sum}\varphi_{i}\left(\phi\right)$
as stated in the beginning of the section.

\subsection{\label{sub:Lemma}Multiplication of paths}

Consider two paths $O_{1}\left(\phi\right),\ O_{2}\left(\phi\right)$
in $SO(N)$ with $\mathbb{Z}_{2}$ indices $\nu_{1},\ \nu_{2}$ respectively.
Then, the path $O\left(\phi\right)=O_{1}\left(\phi\right)O_{2}\left(\phi\right)$
has $\mathbb{Z}_{2}$ index $\nu=\nu_{1}+\nu_{2}$. Indeed, assume
first that one of the paths is contractible (without loss of generality
choose it to be $O_{1}$) while the other one is not. Then there exists
a continuous deformation $O_{1}\left(\phi,t\right)$, with $t\in\left[0,1\right]$
such that $O_{1}\left(\phi,t=0\right)=O_{1}\left(\phi\right)$ and
$O_{1}\left(\phi,t=1\right)=\mathbb{I}$. Then $O\left(\phi,t\right)=O_{1}\left(\phi,t\right)O_{2}\left(\phi\right)$
defines a continuous deformation of $O\left(\phi\right)$ into $O_{2}\left(\phi\right)$
and thus the two must be of the same homotopy class. Next, assume
that both paths are contractible. Then, similarly, one can define
a continuous deformation which takes $O\left(\phi\right)$ into $\mathbb{I}$,
meaning that $O\left(\phi\right)$ is contractible. If both paths
are not contractible, then both of them can be deformed into the same
non-contractible path that we denote by $N\left(\phi\right)$. An
explicit choice of such path can be $\left(\begin{array}{ccccc}
\cos\phi & \sin\phi\\
-\sin\phi & \cos\phi\\
 &  & 1\\
 &  &  & ...\\
 &  &  &  & 1
\end{array}\right)$. Note however that $N\left(\phi\right)^{2}$ is contractible, and
therefore so is $O\left(\phi\right)$.

\subsection{Calculation of the $\mathbb{Z}_{2}$ invariant for a specific model}

We now use the model of a centrosymmetric wire with spin orbit coupling
considered in the main text (Eq. 2). 

\begin{equation}
H_{0}=\xi_{k}+t_{\perp}\sigma_{x}+\lambda_{k}s_{z}\sigma_{z},
\end{equation}
where $\xi_{k}=2t\left(1-\cos k\right)-\mu$, $\lambda_{k}=2\lambda\sin k$
and calculate the $\mathbb{Z}_{2}$ invariant for the cycle in which
the relative phase between the superconductors on the two sides of
the wire vary by $2\pi$ one with respect to the other.

The pairing potential is 
\begin{equation}
H_{\Delta}=\frac{\left(1+\sigma_{z}\right)}{2}\left(1-\beta\right)\Delta\tau_{x}+\frac{\left(1-\sigma_{z}\right)}{2}\Delta\left(\cos\phi\tau_{x}-\sin\phi\tau_{y}\right).
\end{equation}

The phase $\phi$ is varied from $0$ to $2\pi$, and $\beta$ parametrizes
the difference in the magnitude of the gap on the two sides of the
wire. We take $\beta\ll1$, but non zero, to avoid discontinuities
in the eigenstates of the system. 

At the first step, we need to find the negative energy eigenstates
of this BdG Hamiltonian.

First, notice that $s_{z}$ is a good quantum number. Fixing $s_{z}=\pm1$
leaves us with two decoupled copies of a 4$\times$4 Hamiltonian with
spin-orbit coupling $\pm\lambda$ respectively. Since $H$ is both
particle-hole symmetric and inversion symmetric, the bands of $H^{2}$
are doubly degenerate. We thus first diagonalize $H^{2}$ and find
the doubly degenerate subspaces. Note also that the subspaces are
orthogonal to each other since they are eigenstates of $H^{2}$ with
different eigenvalues . Then, we calculate $H$ in each subspace of
$H^{2}$ found previously, and find its eigenstates.

In the $\beta\rightarrow0$ limit: 

\begin{equation}
H^{2}=2\left(\xi_{k}t_{\perp},\ -t_{\perp}\left(\Delta\sin\phi\tau_{x}-\Delta\left(\left(1-\beta\right)-\cos\phi\right)\tau_{y}\right),\ \xi_{k}\lambda_{k}\right)\cdot\vec{\sigma}+\left(\mathrm{diagonal\ part}\right).
\end{equation}

By diagonalizing first the $\tau$ subspace and then the $\sigma$
subspace, we obtain the two doubly degenerate subspaces of $H^{2}$.

The first is spanned by 
\begin{equation}
\Psi_{1}=\left|+u\right\rangle _{\tau}\otimes\left|-\vec{v}_{+}\right\rangle _{\sigma}\ \Psi_{2}=\left|-u\right\rangle _{\tau}\otimes\left|-\vec{v}_{-}\right\rangle _{\sigma}
\end{equation}

where the index $\sigma$/$\tau$ of the states denotes the corresponding
subspace and the eigenvectors are $\left|\pm u\right\rangle =\frac{1}{\sqrt{2}}\left(\begin{array}{c}
\pm1\\
e^{i\gamma}
\end{array}\right)$ with $e^{i\gamma}=\frac{\left(\sin\phi-i\left(\left(1-\beta\right)-\cos\phi\right)\right)}{\sqrt{2\left(1-\beta\right)\left(1-\cos\phi\right)+\beta^{2}}}$
and $\left|-\vec{v}_{\pm}\right\rangle =\left(\begin{array}{c}
-be^{\mp i\alpha}\\
a
\end{array}\right)$ with $a=\frac{1}{\sqrt{2}}\left(1+\frac{\xi_{k}\lambda_{k}}{\sqrt{t^{2}\left(\xi_{k}^{2}+\Delta_{1}^{2}\right)+\xi_{k}^{2}\lambda_{k}^{2}}}\right)^{1/2}$,
$b=\sqrt{1-a^{2}}$, $e^{i\alpha}=\frac{\xi_{k}-i\Delta_{1}}{\left|\xi_{k}-i\Delta_{1}\right|}$.

Note that a non-zero $\beta$ guarantees that $\Delta_{1}\equiv\frac{\Delta}{2}\sqrt{2\left(1-\beta\right)\left(1-\cos\phi\right)+\beta^{2}}>0$
for all $\phi$. Thus, both $e^{i\alpha}$ and $e^{i\gamma}$ are
well defined in the entire parameters space we are interested in. 

The other, orthogonal subspace is spanned by

\begin{equation}
\Psi_{3}=\left|+u\right\rangle _{\tau}\otimes\left|\vec{v}_{+}\right\rangle _{\sigma}\ \Psi_{4}=\left|-u\right\rangle _{\tau}\otimes\left|\vec{v}_{-}\right\rangle _{\sigma}
\end{equation}

where $\left|\vec{v}_{\pm}\right\rangle =\left(\begin{array}{c}
a\\
be^{\pm i\alpha}
\end{array}\right)$ are the states orthogonal to $\left|-\vec{v}_{\pm}\right\rangle $.

Calculating $H$ in the subspace spanned by $\left\{ \Psi_{1},\Psi_{2}\right\} $
we obtain the following Hamiltonian:

\begin{equation}
H_{1,2}=\left[\lambda_{k}-\xi_{k}\left(1-\frac{t^{2}}{r^{2}+\xi_{k}\lambda_{k}}\left[1+\frac{1}{2}\frac{\Delta^{2}}{r^{2}}\beta\right]\right)\right]\rho_{x}-\Delta_{1}\left(1-\frac{t^{2}}{r^{2}+\xi_{k}\lambda_{k}}+\frac{1}{2}\frac{\Delta^{2}}{\Delta_{1}^{2}}\beta\right)\rho_{y}+\frac{\Delta^{2}}{2\Delta_{1}}\sin\phi\rho_{z}\equiv\vec{n}\cdot\vec{\rho},
\end{equation}

where $r^{2}=\sqrt{t^{2}\left(\xi_{k}^{2}+\Delta_{1}^{2}\right)+\xi_{k}^{2}\lambda_{k}^{2}}$
and $\rho_{x,y,z}$ are now Pauli matrices in $\left\{ \Psi_{1},\Psi_{2}\right\} $
subspace.

We now assume that $\xi_{k}>0$ for all $k\in\left[0,\pi\right]$
( i..e $\mu<0$). Otherwise, for the case of $\lambda<0$, it will
be impossible to define single valued eigenstates which are continuous
in the half torus $\phi\in\left[0,2\pi\right],\ k\in\left[0,\pi\right]$.
Note however that as long as $\mu>-t_{\perp}$ we are still in the
topological phase.

For $\lambda>0$ ($s_{z}=+1$), the lower energy eigenstate can be
written simply as $\left|-\vec{n}_{\lambda>0}\right\rangle =\left(\begin{array}{c}
-v_{1}\\
u_{1}e^{i\varphi}
\end{array}\right)$ where $u_{1}=\frac{1}{\sqrt{2}}\left(1+\frac{n_{z}}{\left|n\right|}\right)^{1/2},\ v_{1}=\sqrt{1-u_{1}^{2}},\ e^{i\varphi_{1}}=\frac{n_{x}+in_{y}}{|n_{x}+in_{y}}$.
The phase $\varphi_{1}$ is well defined for the region in parameter
space we are considering.

For $\lambda<0$ ($s_{z}=-1$), to choose a single valued gauge on
the half torus we are interested in one has to rotate $H_{1,2}$ by
$U=e^{i\frac{\pi}{4}\rho_{y}}$. This transforms $H_{1,2}=\vec{n}\cdot\vec{\rho}$
to $\tilde{H}_{1,2}=\vec{\tilde{n}}\cdot\vec{\rho}$ with $\vec{\tilde{n}}=\left(n_{z},\ n_{y},\ -n_{x}\right)$.
The occupied eigenstate of $\tilde{H}_{1,2}$ can be chosen as $\left|-\vec{\tilde{n}}\right\rangle =\left(\begin{array}{c}
-v_{2}e^{-i\varphi_{2}}\\
u_{2}
\end{array}\right)$ where $u_{2}=\frac{1}{\sqrt{2}}\left(1-\frac{n_{x}}{\left|n\right|}\right)^{1/2},\ v_{2}=\sqrt{1-u_{2}^{2}},\ e^{i\varphi_{2}}=\frac{n_{z}+in_{y}}{|n_{z}+in_{y}}$.
This choice of gauge guarantees the eigenstate is single valued. Rotating
the eigenstate back, we obtain $\left|-\vec{n}_{\lambda<0}\right\rangle =e^{i\frac{\pi}{4}\rho_{y}}\left(\begin{array}{c}
-v_{2}e^{-i\varphi_{2}}\\
u_{2}
\end{array}\right)$.

Calculating $H$ also in the subspace spanned by $\left\{ \Psi_{3},\Psi_{4}\right\} $
we obtain

\begin{equation}
H_{3,4}=\left[-\lambda_{k}-\xi_{k}\left(1+\frac{t^{2}}{r^{2}+\xi_{k}\lambda_{k}}\left[1+\frac{1}{2}\frac{\Delta^{2}}{r^{2}}\beta\right]\right)\right]\rho_{x}+\Delta_{1}\left(1+\frac{t^{2}}{r^{2}+\xi_{k}\lambda_{k}}-\frac{1}{2}\frac{\Delta^{2}}{\Delta_{1}^{2}}\beta\right)\rho_{y}+\frac{\Delta^{2}}{2\Delta_{1}}sin\phi\rho_{z}\equiv\vec{n'}\cdot\vec{\rho}
\end{equation}

The eigenstates can be simply taken as $\left|-\vec{n'}\right\rangle =\left(\begin{array}{c}
-v'\\
u'e^{i\varphi'}
\end{array}\right)$ where $u'=\frac{1}{\sqrt{2}}\left(1+\frac{n'_{z}}{\left|n'\right|}\right)^{1/2},\ v'=\sqrt{1-u'^{2}},\ e^{i\varphi'}=\frac{n'_{x}+in'_{y}}{|n'_{x}+in'_{y}}$.
Note that there is no problem with choosing a gauge both for $\lambda>0$
and $\lambda<0$, since $n'_{x}<0$ on the half torus $\phi\in\left[0,2\pi\right],\ k\in\left[0,\pi\right]$.. 

Now we have to write the four negative energy eigenstates we found
in the original basis of $s\otimes\sigma\otimes\tau$, and transform
them to the basis in which particle-hole operator is $\tilde{C}=K$,
by $U=e^{i\frac{\pi}{4}\left(1-s_{y}\tau_{y}\right)}$.

Constructing the matrices $W_{0,\pi}\left(\phi\right)$ and calculating
their eigenvalues numerically, we obtain Fig. \ref{fig:Eigenvalues}.
As one can clearly see, the number of crossings at $\varphi=\pi$
for $k=0$ is three (odd), while the number of crossings for $k=\pi$
is zero (even). The difference corresponds to a non zero $\mathbb{Z}_{2}$
index, and thus the considered cycle is non-trivial.

\begin{figure}[h]
\subfloat[]{\includegraphics[scale=0.5]{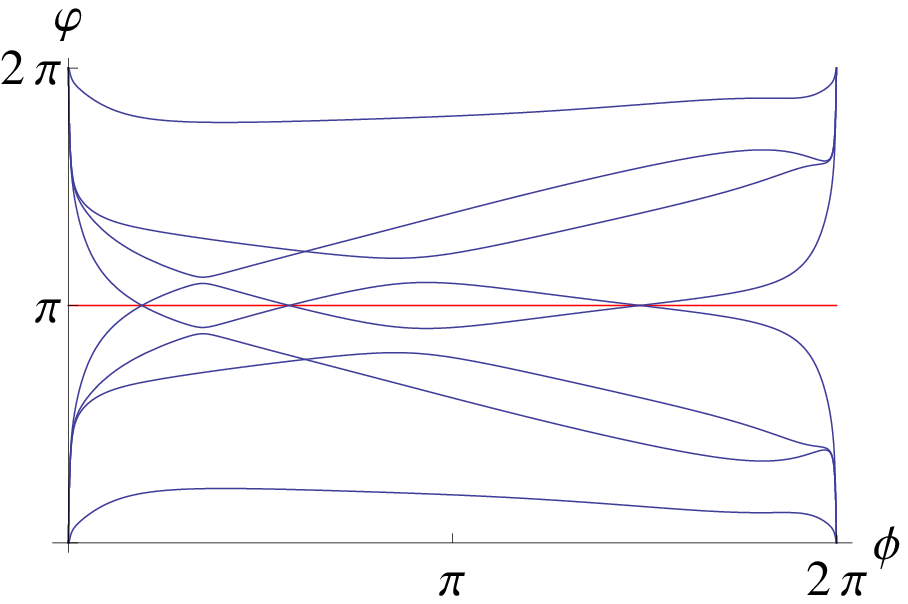}

}\subfloat[]{\includegraphics[scale=0.5]{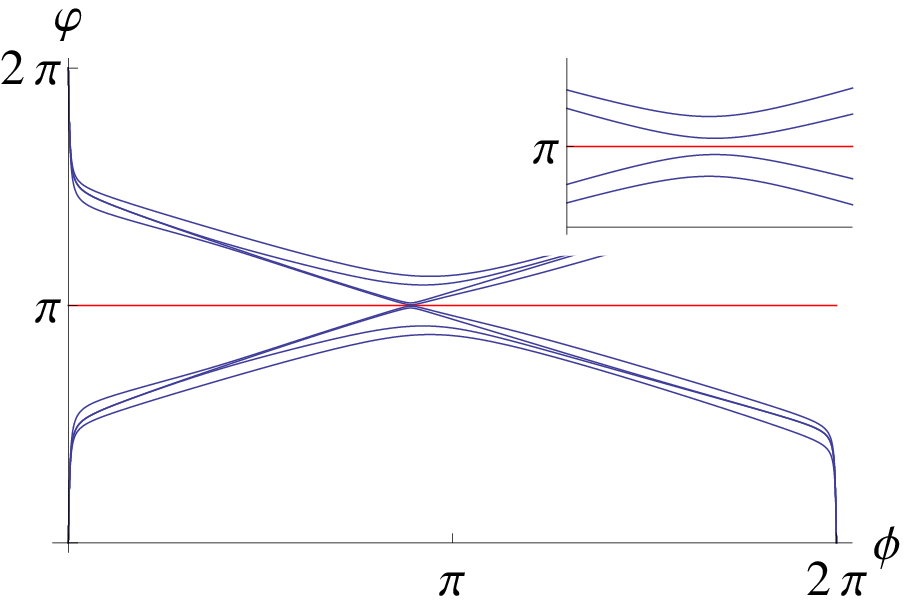}

}

\caption{Eigenvalues of the matrices (a) $W_{k=0}\left(\phi\right)$ and (b)
$W_{k=\pi}\left(\phi\right)$. The lattice model parameters are taken
to be $t_{\perp}=2.5,\ t=1,\ \lambda=1$, and the chemical potential
is set to $\mu=-0.1$. The magnitude of the SC pairing is $\Delta=0.4$,
and the relative difference in the magnitude between the two sides
of the wire is $\beta=0.01$. For $k=0$ there are three crossings
at $\varphi=\pi$ (indicated in red), while for $k=\pi$ the eigenvalues
do not cross the $\varphi=\pi$ line at all. \label{fig:Eigenvalues}}
\end{figure}

\end{widetext}

\end{document}